\documentclass[fleqn,10pt]{temp.iorst} % Document font size and equations flushed left

\definecolor{color1}{RGB}{0,120,0} % Color of the article title and sections
\definecolor{color2}{RGB}{60,0,0} 
\newlength{\tocsep} 
\usepackage{lipsum} % Required to insert dummy text

\definecolor{OliveGreen}{RGB}{0,90,10}

\JournalInfo{Instituto de Estudos Avançados Transdisciplinares}
\Archive{Universidade Federal de Minas Gerais} 
\PaperTitle{The origins of the Malagasy people, some certainties and a few
mysteries} 
\Authors{Maurizio Serva\textsuperscript{1}} % Authors
\affiliation{\textsuperscript{1}\textit{Dipartimento di Ingegneria e 
Scienze dell'Informazione e Matematica, Universit\`{a} dell'Aquila, 
I-67010 L'Aquila, Italy}}  % Author affiliation

\newcommand{\keywordname}{Keywords} % Defines the keywords heading name

\Abstract{
The Malagasy language belongs to the Greater Barito East group of the
Austronesian family, the language most closely connected to Malagasy dialects 
is Maanyan (Kalimantan), but 
Malay as well other Indonesian and Philippine languages are also related.
The African contribution is very high in the Malagasy genetic
make-up (about 50\%) but negligible in the language.

Because of the linguistic link, it is widely accepted that 
the island was settled by Indonesian 
sailors after a maritime trek but date and place of landing
are still debated. 
The 50\% Indonesian genetic contribution to present Malagasy 
points in a different direction then Maanyan for the Asian ancestry, 
therefore, the ethnic composition of the Austronesian settlers
is also still debated.

In this talk I mainly review the joint research of
{\color{OliveGreen} 
Filippo Petroni, Dima Volchenkov,  S\"oren Wichmann and myself}
which tries to shed 
new light on these problems. The key point is the application of 
a new quantitative methodology
which is able to find out the kinship relations among languages
(or dialects). New techniques are also introduced in
order to extract the maximum information from these relations concerning 
time and space patterns.}
\begin{document}

\flushbottom % Makes all text pages the same height

\maketitle % Print the title and abstract box

%\tableofcontents % Print the contents section

\thispagestyle{empty} % Removes page numbering from the first page

\renewcommand{\thefootnote}{} 
\bigskip
\footnote{\normalsize \color{OliveGreen} This text reports the Grande Confer\^encia 
{\it As origens do povo malgaxe, algumas certezas e v\'arios mist\'erios}
given at the  Universidade Federal de Minas Gerais, 
Instituto de Estudos Avan\c{c}ados Transdisciplinares, 
(Belo Horizonte, 11 August 2016).}

%----------------------------------------------------------------------------
%	ARTICLE CONTENTS
%----------------------------------------------------------------------------

\section{Introduction}

\begin{figure}
 \includegraphics[width=3.0truein,height=4.6truein,angle=0]{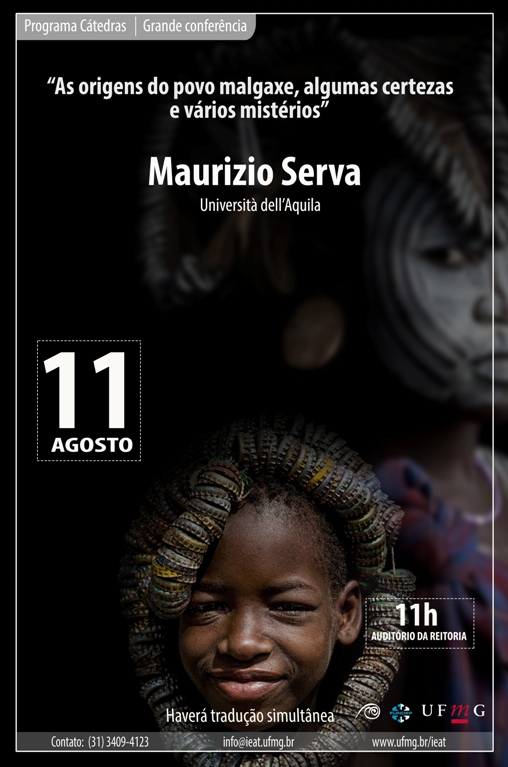}
\end{figure}

\begin{figure}
 \includegraphics[width=3.3truein,height=2.5truein,angle=0]{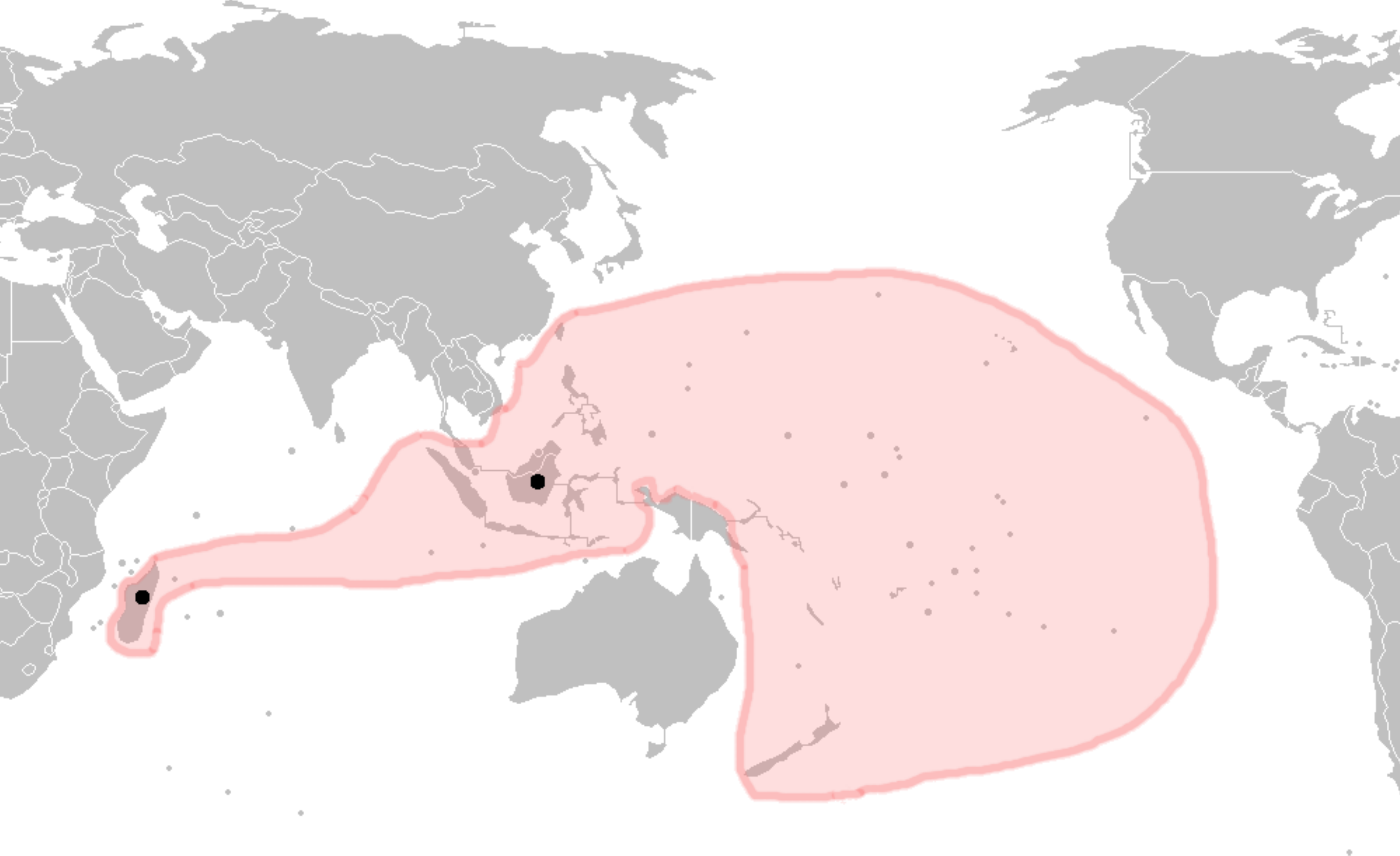}
 \caption{The distribution of Austronesian languages. The black spots 
indicate Madagascar and Kalimantan, where Maanyan, the closest relative of Malagasy, 
is spoken.}
\label{Fig_1}
\end{figure}

The Austronesian expansion, which very likely started from Taiwan or 
from the south of China
\cite{Gray:2000, Hurles:2005}, is probably the most spectacular event 
of maritime colonization in human history as it can be appreciated in Fig.~\ref{Fig_1}.

The Malagasy language (as well as all its dialects) belongs to 
the Austronesian linguistic family, as was suggested already in 
\cite{Houtman:1603} and later firmly established in 
\cite{Tuuk:1864}.
Much more recently, Dahl \cite{Dahl:1951} pointed out a particularly 
close relationship  between Malagasy and Maanyan of south-east Kalimantan, 
which share about 45\% their basic vocabulary \cite{Dyen:1953}. 
But Malagasy also bears similarities to languages in Sulawesi, Malaysia, Sumatra 
and Philippines, including loanwords from Malay, Javanese, 
and one (or more) language(s) of southern Sulawesi \cite{Adelaar:2009}.

The genetic make-up of Malagasy people exhibits almost equal proportions of 
African and Indonesian heritage \cite{Hurles:2005}. Nevertheless
its Bantu component in the vocabulary seems to be very limited and mostly concerns
faunal names \cite{Blench and Walsh:2009}.

The history of Madagascar peopling and settlement 
is subject to alternative interpretations among scholars. 
It seems that Indonesian sailors reached Madagascar by a maritime trek at a 
time between one to two thousand years ago (the exact time and the place of landing
are still debated) but until recently 
it was not clear whether there were multiple settlements or just a single one. 

This last question was answered in \cite{Cox:2012} were it was shown
that Madagascar was settled by a very small group of women (approx. 30). 
This highly restricted founding population suggests that Madagascar
was settled through a single, perhaps even unintended, transoceanic crossing.

Additional questions are raised by the fact that the Maanyan speakers,
which live along the rivers of Kalimantan, have not the 
necessary skills for long-distance maritime navigation. 
Moreover, recent research on DNA, while pointing to south-east Kalimantan and Sulawesi
for the Indonesian ancestry,
firmly rejects a direct genetic link between Malagasy and Maanyan. 
\cite{Kusuma:2015,Kusuma:2016}.
\bigskip

In this talk I review our results which give information about the
following points: 

%\begin{enumerate}
\begin{enumerate}[noitemsep] 
% [noitemsep] removes whitespace between the items for a compact look

\item[$i$)] the historical configuration of Malagasy dialects,

\item [$ii$)] when the migration to Madagascar took place,

\item [$iii$)] how Malagasy is related to other Austronesian languages,

\item [$iv$)] where the original settlement of the Malagasy people took place.

\end{enumerate}

Our research addresses these four problems through the application of 
new quantitative methodologies inspired by, but nevertheless different 
from, classical lexicostatistics and glottochronology 
\cite{Serva:2008, Holman:2008, Petroni:2008, Bakker:2009}.

The data, collected  during the beginning of 2010 by one of the authors 
(M.S.), consist of 200-item 
Swadesh word lists for 23 dialects of Malagasy from all areas of the island. 
A practical orthography which corresponds to the orthographic conventions 
of standard Malagasy has been used.
Most of the informants were able to write the words directly 
using these conventions, while a few of them benefited from the help of one or 
more fellow townsmen. 
A cross-checking of each dialect list was done by eliciting data 
separately from two different consultants.
Details about the speakers who furnished the data are in \cite{Serva:2012a}
while the dataset can be found in \cite{Serva:2011}, 
This dataset probably represents the largest collection available of comparative Swadesh lists for Malagasy (see Fig.~\ref{Fig_4} for the locations). 

While there are linguistic as well as geographical and temporal dimensions to 
the issues addressed in this talk, all strands of the investigation are 
rooted in an automated comparison of words.
Our automated method (see Appendix A for details) works as follows:
for any language we write down a Swadesh list, then 
we compare words with same meaning belonging to
different languages only considering orthographic differences.
This approach is motivated by the analogy with genetics: 
the vocabulary has the role of DNA and the comparison is simply made by 
measuring the differences between the DNA of the two languages.
There are various advantages: 
the first is that, at variance with previous methods, 
it avoids subjectivity, the second is
that results can be replicated by other scholars assuming that the
database is the same, the third is that it is not requested a specific 
expertize in linguistic, and the last, but surely not the least, is
that it allows for a rapid comparison of a very large number of languages
(or dialects).

The first use of the pairwise distances is to derive a classification of the dialects.
For this purpose we adopt a multiple strategy in order to extract a maximum of information
from the set of pairwise distances. 
We first obtain a tree representation of the set by using two different standard 
phylogenetic algorithms,
then we adopt a strategy (SCA) which, analogously to a principal components approach,
represents the set in terms of geometrical relations. 
The SCA analysis also provides the tool for a dating of the landing of Malagasy ancestors
on the island.
The landing area is established assuming that a linguistic homeland is the area
exhibiting the maximum of current linguistic diversity. Diversity is measured by
comparing lexical and geographical distances. Finally, we perform a comparison 
of all variants with some other Austronesian languages, in particular with Malay 
and Maanyan.

For the purpose of comparison of Malagasy variants 
with other Austronesian languages we draw upon {\it The Austronesian 
Basic Vocabulary Database.} \cite{Greenhill:2009}. 
Since the wordlists in this database do not always 
contain all the 200 items of our Swadesh lists they are supplemented 
by various sources (including the database of the Automated Similarity Judgment 
Program (ASJP)\cite{Wichmann:2017}) and by author's interviews.

\section{Method}

Our strategies \cite{Serva:2008, Holman:2008, Petroni:2008, Bakker:2009} 
are based on a lexical comparison 
of languages by means of an automated measure of distance 
between pairs of words with same meaning contained in Swadesh lists.
The use of Swadesh lists \cite{Swadesh:1952} in lexicostatistics has been popular for more than half a century.  
They are lists of words associated with 
the same $M$ meanings,
(the original choice of Swadesh was $M=200$) which tend (1) to be found in all languages, (2) to be relatively stable, 
(3) to not frequently be borrowed, and (4) to be basic rather than derived concepts. It has to be stressed that 
these are tendencies and that convenience play a great part in the way the list of concepts was put together. 
Still, it has become standard, and here we simply follow Swadesh and the many other scholars who have applied it. 
Comparing the two lists corresponding to a pair of languages
it is possible to determine the percentage of shared {\it cognates}
which is a measure of their lexical distance.
A recent example of the use of Swadesh lists and cognates counting to 
construct language trees are the studies of Gray and Atkinson \cite{Gray:2003} 
and Gray and Jordan \cite{Gray:2000}.

The idea of measuring relationships among languages
using vocabulary is much older than lexicostatistics
and it seems to have its roots in the work of 
the French explorer Dumont D'Urville.
He collected comparative word lists during his voyages aboard the Astrolabe 
from 1826 to 1829 and, in his work about 
the geographical division of the Pacific \cite{D'Urville:1832},
he proposed a method to measure the degree of relation 
among languages.
He used a core vocabulary of 115 terms, then he assigned a distance 
from 0 to 1 to any pair of words with the 
same meaning and finally he was able to determine the degree of
relation between any pair of languages.

Our results are obtained through a specific version of the so-called 
Levenshtein or 'edit' 
distance (henceforth LD) \cite{Levenshtein:1966}. 
The version we use here was introduced 
by \cite{Serva:2008, Petroni:2008} and consists of the following 
procedure. Words referring to the same concept for a 
given pair of dialects are compared 
with a view to how easily the word in dialect A is transformed into the corresponding word 
in dialect B. Steps allowed in the transformations are: 
insertions, deletions, and 
substitutions. The LD is then calculated as the minimal 
number of such steps required 
to completely transform one word into the other. 
Calculating the distance measure we use 
(the 'normalized Levenshtein distance', or LDN),
requires one more operation: the 'raw LD' is 
divided by the length (in terms of segments) of the longer of the two words compared. 
This operation produces LDN values between 0 and 1 and takes into account 
variable word lengths: if one or both of the words compared happen to be 
relatively long, 
the LD is prone to be higher than if they both happen to be short, 
so without the 
normalization the distance values would not be comparable. 
Finally we average the LDN's for 
all 200 pairs of words compared to 
obtain a distance value characterizing the overall difference 
between a pair of dialects 
(see Appendix A for a compact mathematical definition and a table with all distances).

Thus, the Levenshtein distance is sensitive to both lexical replacement and phonological 
change and therefore differs from the cognate counting 
procedure of classical lexicostatistics even if the results are usually roughly equivalent.

If a family of languages is considered, all the information is encoded 
in a matrix whose entries are the pairwise lexical distances. 
But information about the total relationship among the 
languages is not manifest and it has to be extracted.  
The ubiquitous approach to this problem
is to transform the matrix information in a phylogenetic tree.

Nevertheless, in this transformation, part of the information may be lost
because transfer among languages is not exclusively vertical
(as in mtDNA transmission from mother to child)
but it also can be horizontal (borrowings and, in extreme cases, creolization).
Another approach is the geometric one \cite{Blanchard:2010a, Blanchard:2010b} 
that results
from Structural Component Analysis (SCA) that we have recently proposed.
This approach encodes the matrix information into the positions of 
the languages in a $n$-dimensional space. 
For large $n$ one recovers all the matrix content, but a low dimensionality, 
typically $n$=2 or $n$=3, is sufficient to grasp all the relevant information.
The results presented in this talk mostly rely to a direct investigation of the 
entries of the matrix and to simple averages over them.

\section{Phylogenetic trees and geography}

The number of Malagasy dialects we consider is $N$=23,
therefore, the output of our method, when applied only
to these variants is a 
matrix with $N(N-1)/2=253$ non-trivial entries
representing all the possible lexical distances among dialects.
This matrix is explicitly shown in Appendix A.

The information concerning the vertical transmission of vocabulary
from proto-Malagasy to the contemporary dialects can be extracted by
a phylogenetic approach. There are various possible choices for the 
algorithm for the reconstruction of the family tree.
The two algorithms used are Neighbor-Joining (NJ) \cite{Saitou:1987} and the
Unweighted Pair Group Method (UPGMA) \cite{Sokal:1958}. 
The main theoretical difference between the algorithms is that UPGMA assumes that 
evolutionary rates are the same on all branches of the tree, while NJ allows differences 
in evolutionary rates. The question of which method is better at inferring the phylogeny 
has been studied by running various simulations where the true phylogeny is known. 
Most of these studies were in biology but at least one \cite{Barbancon:2006} specifically 
tried to emulate linguistic data. Most of the studies (starting with \cite{Saitou:1987} 
and including \cite{Barbancon:2006}) found that NJ usually came closer to the 
true phylogeny. 
Since in our case, the relations among dialects are not necessarily
tree-like, it is desirable to test the different methods against empirical 
linguistic data, which is mainly why trees derived by means of both methods are presented here. 

\begin{figure}
 \includegraphics[width=3.5truein,height=3.46truein,angle=0]{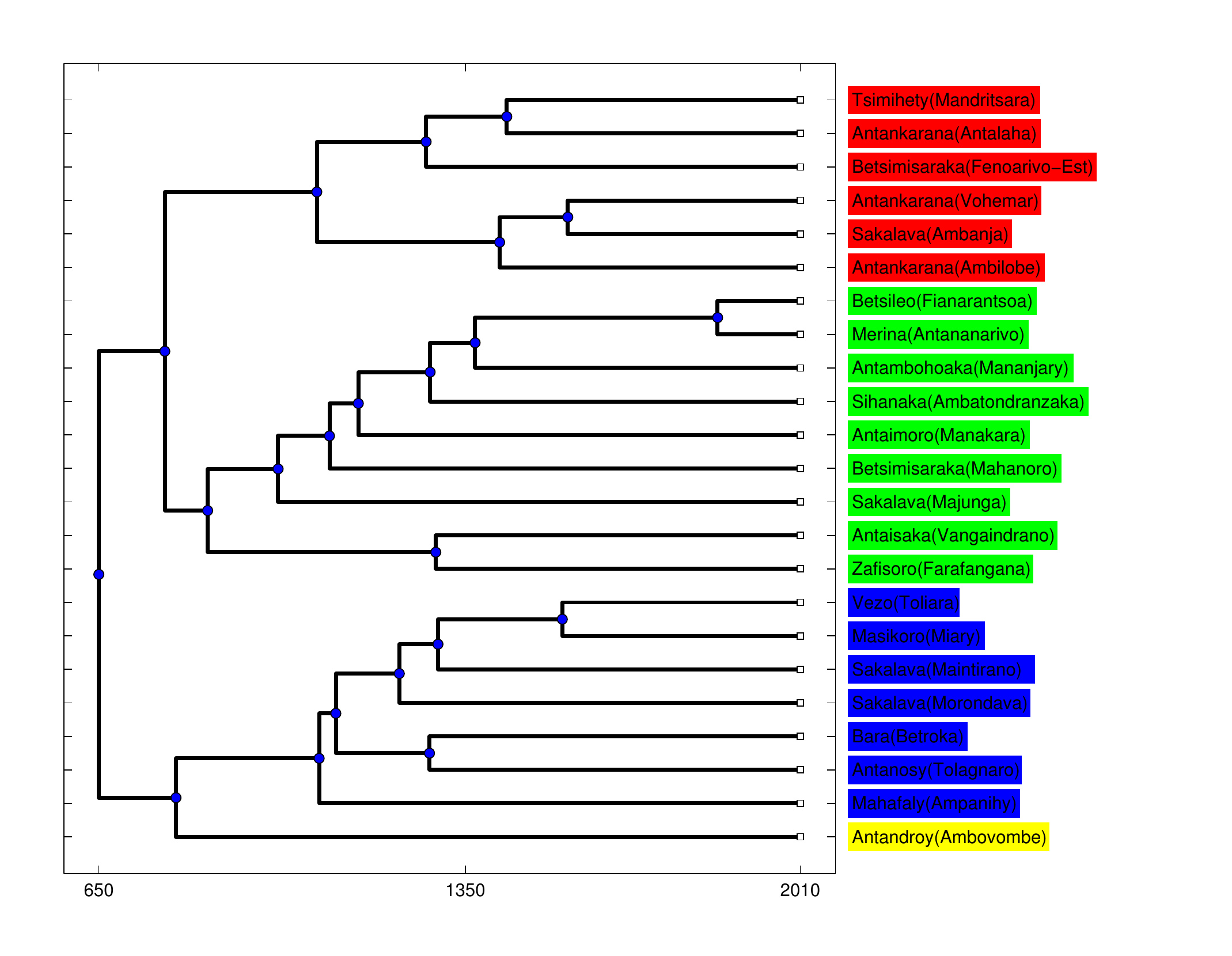}
 \caption{Phylogenetic tree of 23 Malagasy dialects realized by
Unweighted Pair Group Method Average (UPGMA). 
In this figure the name of the dialect is followed by the name of the town were 
it was collected. The phylogenetic tree shows a partition of the Malagasy 
dialects into four main groups indicated by different colors.}
\label{Fig_2}
\end{figure}

\begin{figure}
\includegraphics[width=3.3truein,height=3.3truein,angle=0]{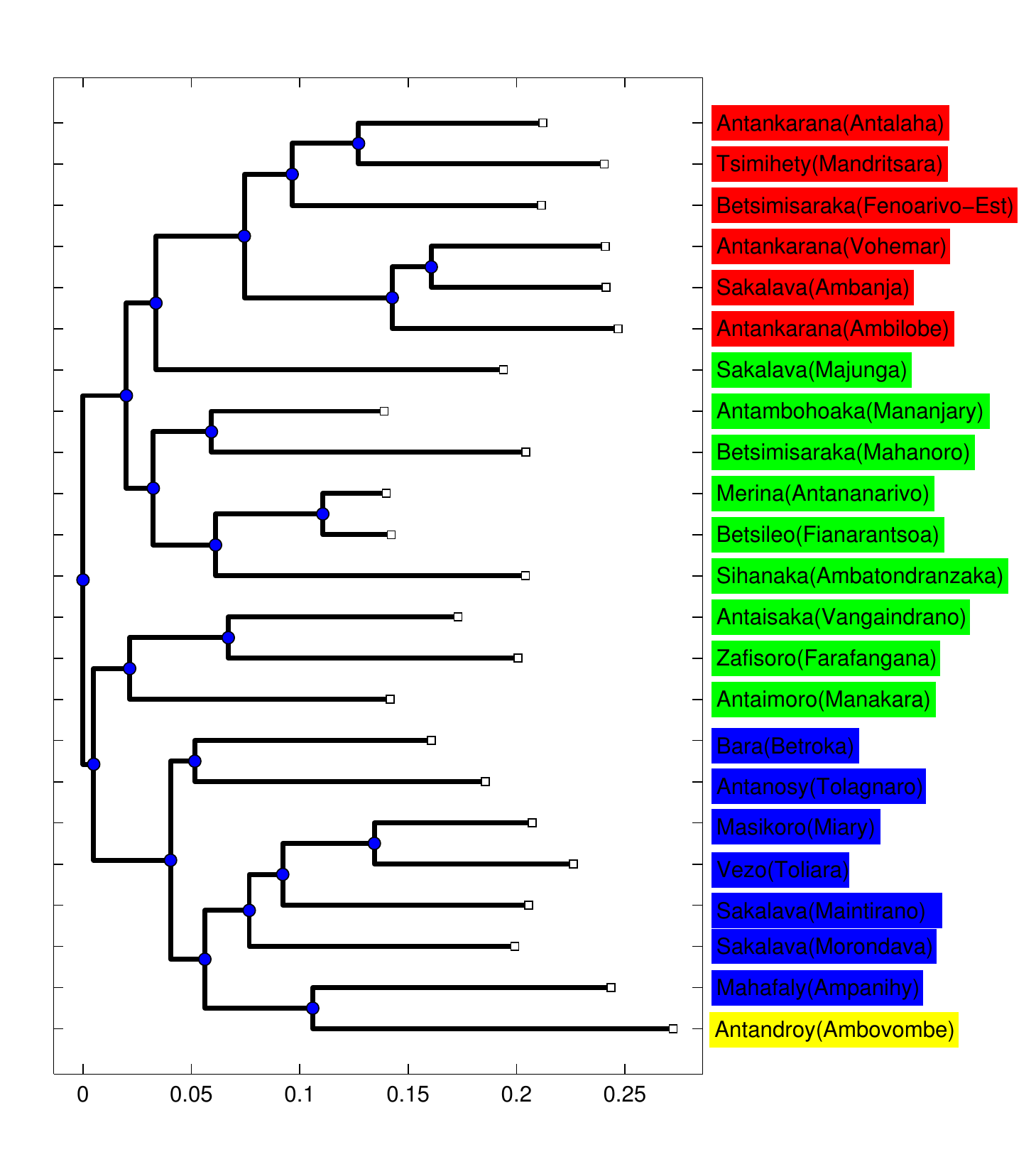}
\caption{NJ tree for 23 Malagasy dialects. Colors 
compare with the UPGMA tree in  Fig. \ref{Fig_2}.
The graph confirms the main center-north-east vs. south-west division.
The main difference is that three dialects at the 
linguistic border are grouped differently. Colors allows for a rapid comparison.}
\label{Fig_3}
\end{figure}

The input data for the UPGMA tree are 
the pairwise separation times obtained from lexical distances by a rule
\cite{Serva:2008} which is a simple generalization of
the fundamental formula of glottochronology.
The absolute time-scale is calibrated by the results of the SCA analysis (see below), 
which indicate a separation date of 650 CE.
While the scale below the UPGMA tree (Fig.~\ref{Fig_2}) refers to separation times, 
the scale below the NJ tree (Fig.~\ref{Fig_3}) simply shows lexical distance from the root.
The LDN distance between two language variants is roughly equal to the sum of their
lexical distance from their closest common node.

Since UPGMA assumes equal evolutionary rates, the ends 
of all the branches line up on the right side of the UPGMA tree. The assumption of equal 
rates also determines the root of the tree on the left side. 
NJ allows unequal rates, so the ends of the branches do not all line up on the NJ tree. 
The extent to which they fail to line up indicates how variable the rates are. 
The tree is rooted by the midpoint (the point in the network in between the two most 
distant dialects) but we also checked that the same result is obtained 
following the standard strategy of adding an out-group.

\begin{figure}
 \includegraphics[width=4.5truein,height=3.5truein,angle=0]{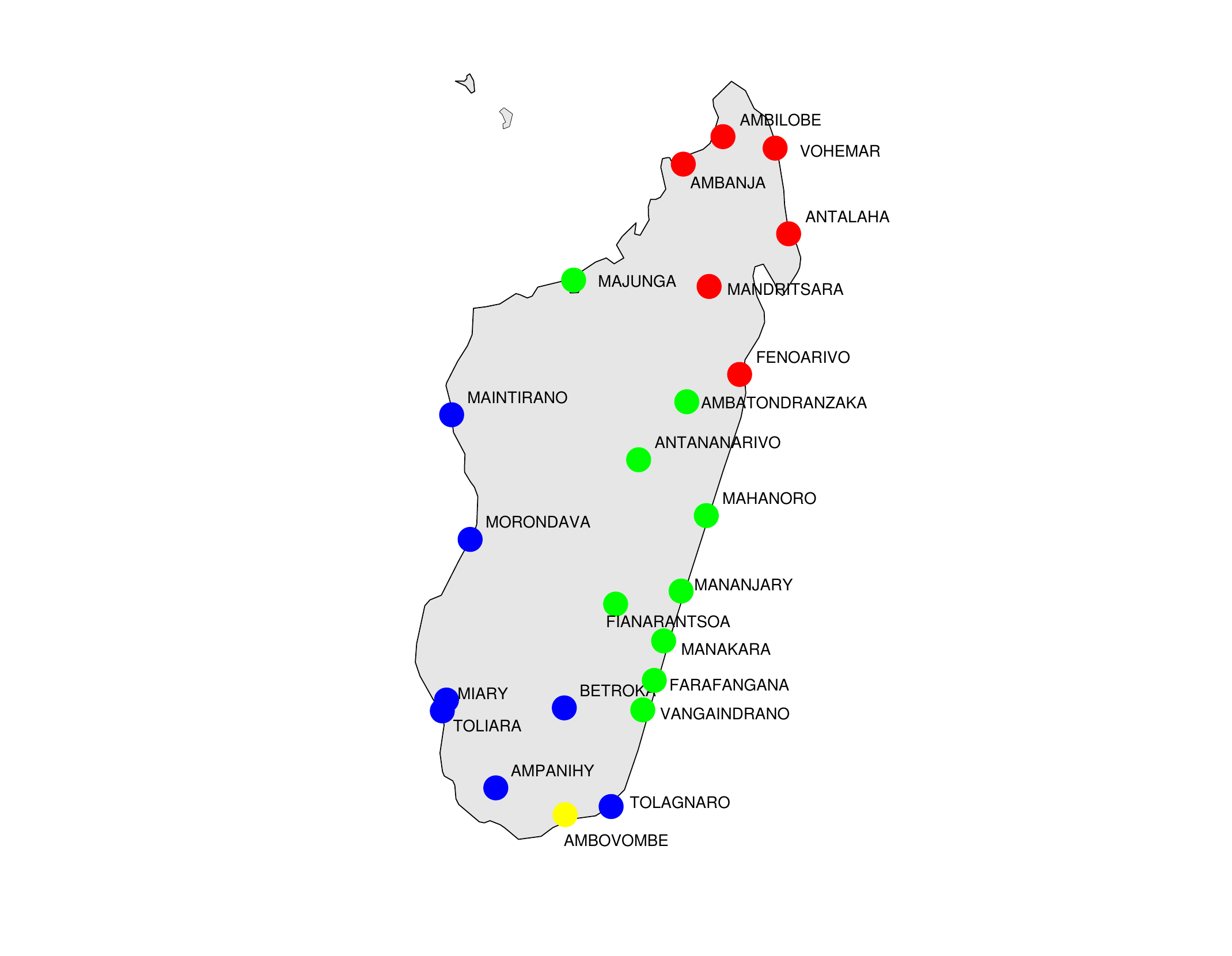}
 \caption{Geography of Malagasy dialects.  The locations of the 23 dialects 
are indicated with the same colors of Fig. \ref{Fig_2}.
Any dialect is identified by the the name of the town where 
it was collected.}
\label{Fig_4}
\end{figure}

There is a good fit
between the geographical position of dialects (see Fig. \ref{Fig_4}) 
and their position
both in the UPGMA (see Fig. \ref{Fig_2}) and NJ trees (see Fig. \ref{Fig_3}).
In both trees the dialects are divided into two main groups (colored blue 
and yellow vs. red and green in Fig.~\ref{Fig_2}).

Given the consensus between the two methods, the result regarding the 
basic split can be considered solid. 
Geographically the division corresponds to a border 
running from the south-east to the north-west of the island, as shown in 
Fig.~\ref{Fig_4}
where the UPGMA and NJ main separation lines are drawn.
A major difference concerns the Vangaindrano, Farafangana and Manakara dialects, 
which have shifting allegiances with respect to the two main groups under the different analyses. Additionally, 
there are minor differences in the way that the two main groups are configured 
internally. Most strikingly, we observe that in the UPGMA tree
Majunga (a.k.a. Mahajanga) is grouped with the central dialects while in the NJ tree it is grouped
with the northern ones. This indeterminacy would seem to relate 
to the fact that the town 
of Majunga is at the geographical border of the two regions.

Another difference is that in the UPGMA tree the Ambovombe variant 
of the dialect traditionally called Antandroy is quite isolated, whereas in the NJ tree 
Ambovombe and the Ampanihy variant of Mahafaly group together. 
Since the UPGMA algorithm is a strict bottom-up approach to the construction of a phylogeny, 
where the closest taxa a joined first, it will tend to treat the overall most deviant variant 
last. In contrast, the NJ algorithm privileges pairwise similarities. This explains the 
differential placement of Ambovombe in the two trees. 

The length of the branch leading to the node that joins Ambovombe and Ampanihy in the NJ 
tree shows that these two variants have quite a lot of similarities but 
in the UPGMA method these similarities in a sense 'drown' in 
the differences that set Ambovombe off from other Malagasy variants {\it as a whole}. 

The phylogenetic trees interestingly shows a main 
partition of Malagasy dialects in two main branches (east-center-north 
and south-west) at variance with a previous study which gave a different
partitioning \cite{Verin:1969} isolating northern dialects (indeed,
results in \cite{Verin:1969} coincide with ours if a modern phylogenetic 
algorithm is applied to their data, see 
\cite{Serva:2012a} for a discussion of this point.)

\section{Structural Component Analysis}

Although tree diagrams have become ubiquitous in representations of language taxonomies, 
they fail to reveal the full complexity of affinities among languages.
The reason is that the simple relation of ancestry, which is the single principle behind a 
branching family tree 
model, cannot grasp the complex social, cultural and political factors molding  
the evolution of languages \cite{Heggarty:2006}. 
Since all dialects within a group interact with each other 
and with the languages of other families in 'real time', it is obvious that any historical 
development in languages cannot be described only in terms of pair-wise interactions, 
but reflects a genuine higher order influence, which can best be assessed by Structural 
Component Analysis (SCA). This is a powerful tool which represents the relationships 
among different languages in a language family geometrically, in terms of distances 
and angles, as in the Euclidean geometry of everyday intuition. Being a version 
of the kernel PCA method \cite{Schoelkopf:1998}, it generalizes PCA to cases where we are 
interested in principal components obtained by taking all higher-order correlations 
between data instances. It has so far been tested through the construction of language 
taxonomies for fifty major languages of the Indo-European and Austronesian language 
families \cite{Blanchard:2010a}. The details of the SCA method are given in the Appendix B. 

In Fig.~\ref{Fig_5} we show the three-dimensional geometric representation of 23 
dialects of the Malagasy language and the Maanyan language, which is closely related to 
Malagasy.
The three-dimensional space is spanned by 
the three major data traits ($\{q_2,q_3,q_4\}$, see Appendix B for details) 
detected in the matrix of linguistic LDN distances.

\begin{figure}
 \includegraphics[width=3.25truein,height=3.25truein,angle=0]{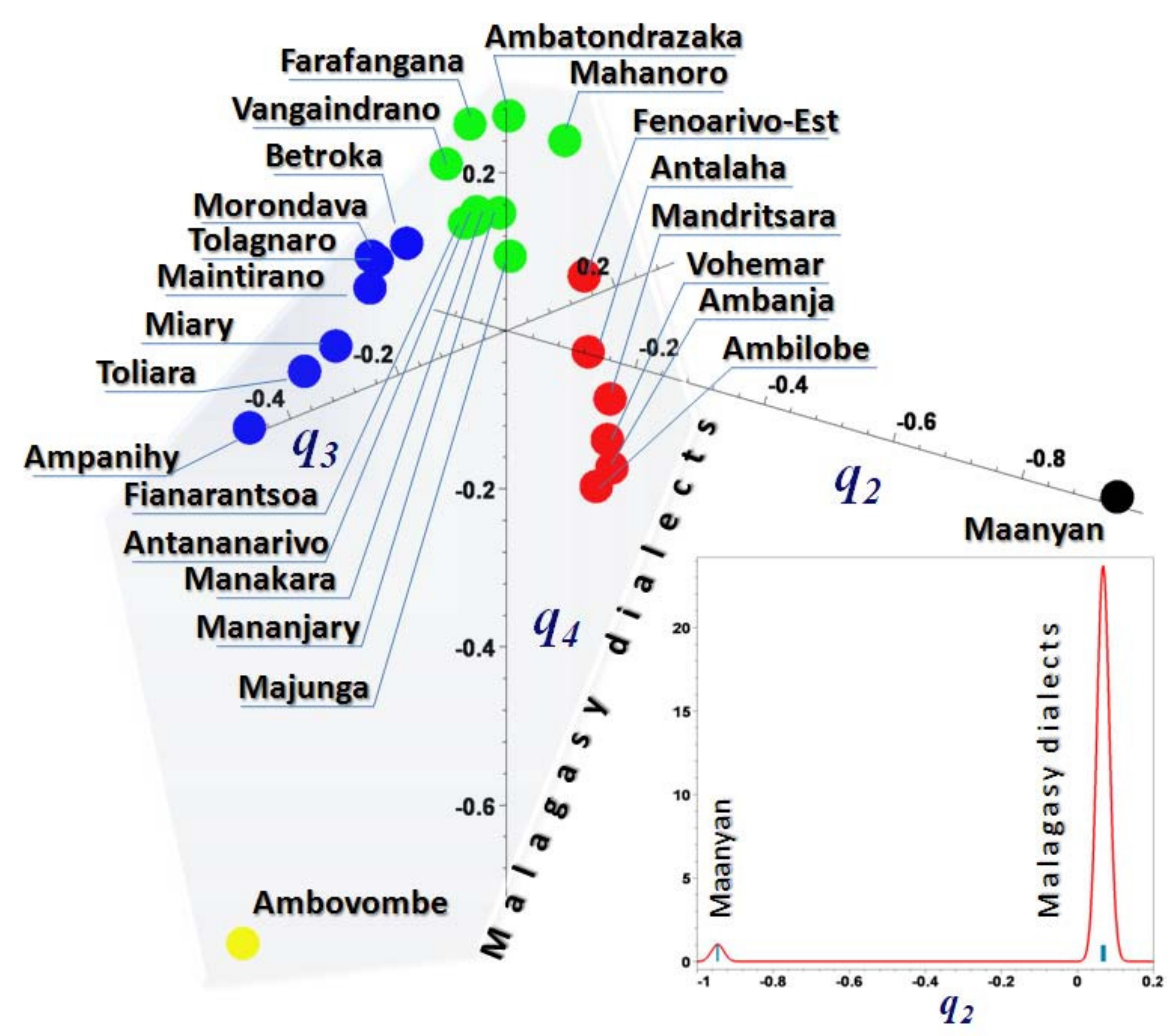}
 \caption{The three-dimensional geometric representation of the Malagasy dialects
and the Maanyan language in the space of major data traits ($q_2, q_3, q_4$) 
shows a remarkable geographic patterning separating the northern (red) and the southern 
(blue) dialect groups, which fork from the central part of the island 
(the dialects spoken in the central part are colored green, while Antandroy is yellow).
The kernel density estimate of the distribution
of the $q_2$ coordinates, together with the absolute data frequencies,
indicate that all Malagasy dialects belong to a single plane 
orthogonal to the data trait of the Maanyan language ($q_2$).}
\label{Fig_5}
\end{figure}

\begin{figure}
 \includegraphics[width=3.5truein,height=3.45truein,angle=0]{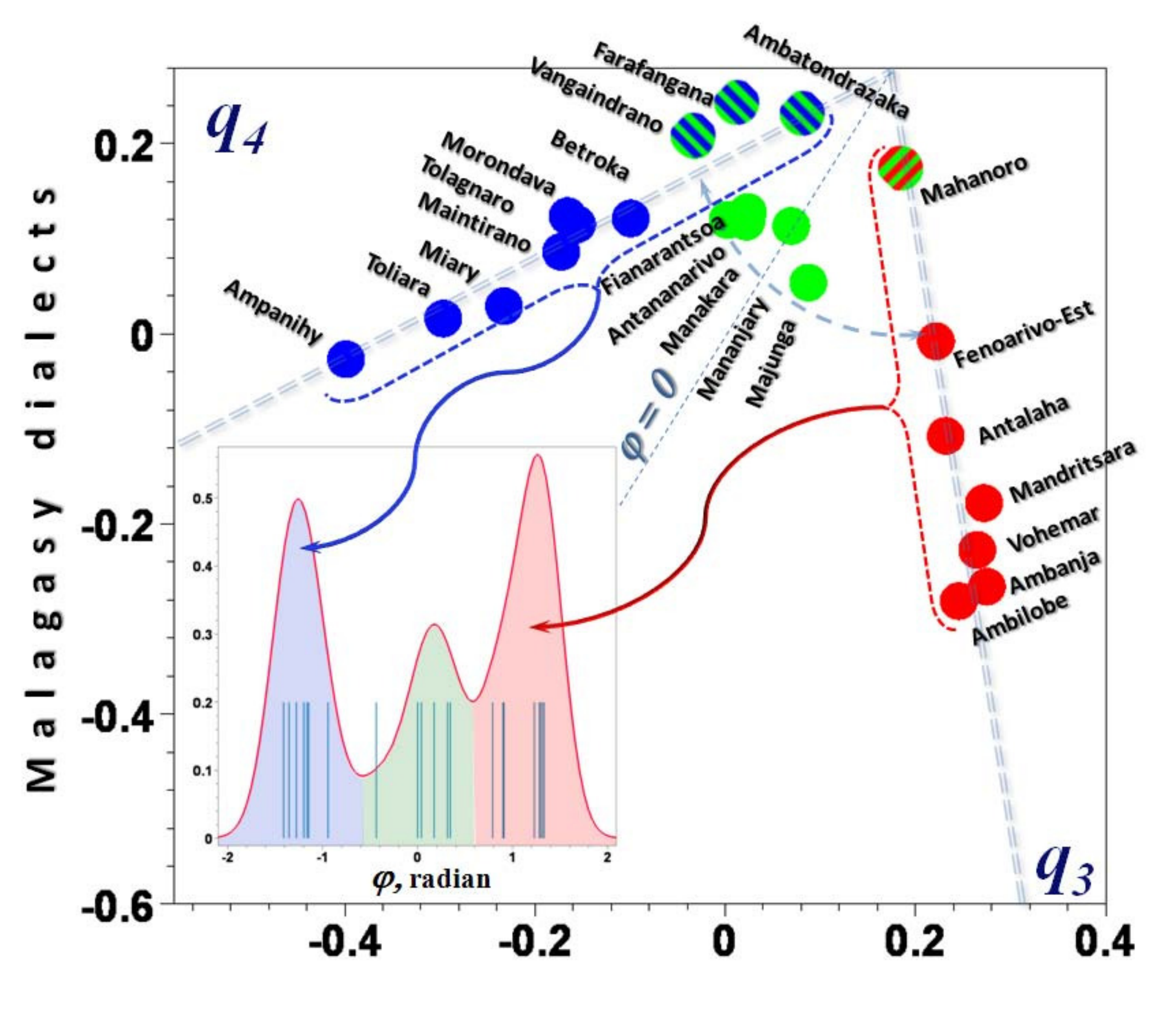}
 \caption{The plane of Malagasy dialects ($q_3,q_4$); Antandroy (Ambovombe) is excluded.
The kernel density estimate of the distribution over azimuth angles, together with the 
absolute data frequencies, allows the rest of Malagasy dialects to be classified 
into the three groups: north (red), south-west (blue), and center (green).}
\label{Fig_6}
\end{figure}

The clear geographic patterning is perhaps the most remarkable aspect of the geometric 
representation. The structural components reveal themselves in Fig.~\ref{Fig_6} as two 
well-separated spines representing both the northern (red) and the southern (blue) 
dialects of entire language.

\begin{figure}
 \includegraphics[width=3.3truein,height=3.0truein,angle=0]{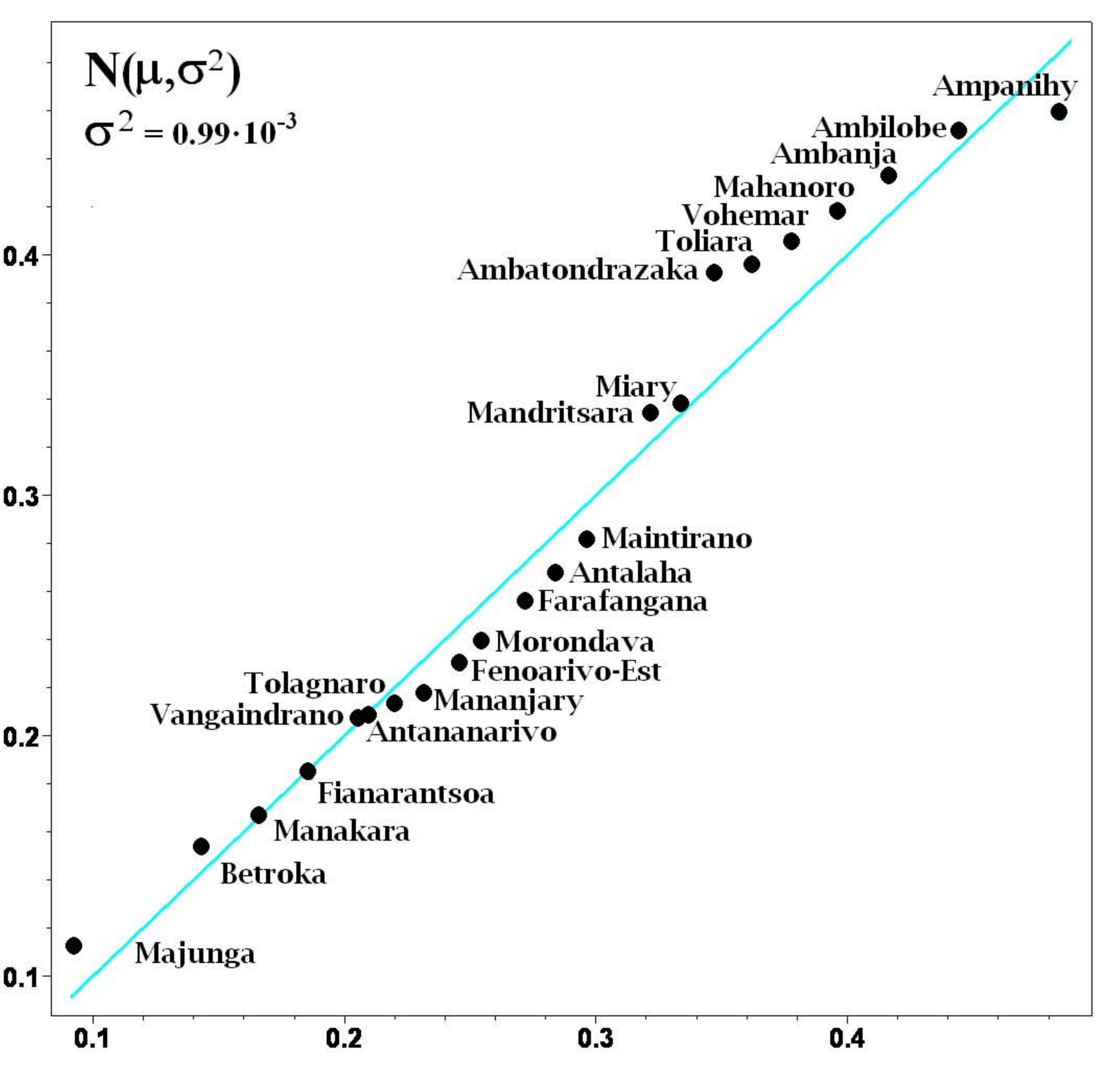}
 \caption{The radial coordinates are ranked and then plotted
against their expected values under normality. Departures from linearity, 
which signify departures from normality, is minimal.}
\label{Fig_7}
\end{figure}

It is remarkable that all 
Malagasy dialects belong to a single plane orthogonal to the data trait of the Maanyan 
language ($q_2$). 
The plane of Malagasy dialects is attested by the sharp distribution of 
the language points in Cartesian coordinates along the data trait $q_2.$ 
This color point of Malagasy dialects over their common plane is shown in Fig.~\ref{Fig_7}
where a reference azimuth angle $\varphi$ is introduced in order to
underline the evident symmetry. 
It is important to mention that although the language point of Antandroy (Ambovombe) is 
located on the same plane as the rest of Malagasy dialects, it is situated far away from 
them and obviously belongs to neither of the dialect branches and for this reason is not 
reported in next Fig.~\ref{Fig_7}.
This clear SCA isolation of Antrandroy is compatible with its position in the tree  
in  Fig.~\ref{Fig_2}.

The distribution of language points supports the main conclusion following from 
the UPGMA and NJ methods (Figs.~\ref{Fig_1}-\ref{Fig_2}) of a division of the 
main group of Malagasy dialects into three groups: north (red), south-west (blue) 
and center (green). These clusters are clearly evident from the representation shown 
in Fig.~\ref{Fig_7}. However, with respect to the classification of some individual 
dialects the SCA method differs from the UPGMA and NJ results. 
Since their azimuthal coordinates better fit the general trend of the southern group, 
the Vangaindrano,  Farafangana, and  Ambatontrazaka dialects spoken in the central part 
of the island are now grouped with 
the southern dialects (blue) rather than the central ones . 
Similarly, in accordance with the representation shown in Fig.~\ref{Fig_7}, the Mahanoro 
dialect is now classified in the northern group (red), since it is best fitted to the 
northern group azimuth angle. The remaining five dialects of the central group (green colored) 
are characterized by the azimuth angles close to a bisector ($\varphi=0$).

\section{A date for landing}

The radial coordinate of a dialect is simply the distance of
its representative point from the origin of coordinates in Fig.~\ref{Fig_5}.
It can be verified that the position of Malagasy dialects along the radial 
direction is remarkably heterogeneous indicating that the rates of change in the 
Swadesh vocabulary was anything but constant. 
  
The radial coordinates have been ranked and then plotted in Fig.~\ref{Fig_7} 
against their expected values under normality, such that departures from linearity 
signify departures from normality.
The dialect points in Fig.~\ref{Fig_7} show very good agreement with 
univariate normality with the value of variance $\sigma^2 = 0.99\times 10^{-3}$ 
which results from the best fit of the data. This normal behavior
can be justified by the hypothesis that the dialect vocabularies are the result of a
gradual and cumulative process into which many small, independent innovations 
have emerged and contributed additively.

In the SCA method, which is based on the 
statistical evaluation of differences among the items of the Swadesh list, 
a complex nexus 
of processes behind the emergence and differentiation of 
dialects is described by the single degree of freedom (as another degree of freedom, 
the azimuth angle, is fixed by the dialect group) along the radial direction
\cite{Blanchard:2010b}.
 
The univariate normal distribution (Fig.~\ref{Fig_7}) 
implies a homogeneous diffusion time evolution in one dimension,  under which variance 
$\sigma^2\propto t$  grows linearly with time. The locations of dialect 
points could not 
be distributed normally if in the long run the value of variance 
$\sigma^2$ did not grow 
with time at an approximately constant rate. We stress that the constant rate of increase
in the variance of radial positions of languages  in the geometrical representation 
(Fig.~\ref{Fig_5}) has nothing to do with the traditional glottochronological assumption 
about the constant replacement rate of cognates assumed by the UPGMA method. 

It is also important to mention that the value of variance $\sigma^2= 0.99\times 10^{-3}$ 
calculated for 
the Malagasy dialects does not correspond to physical time but rather gives a statistically 
consistent estimate of age for the group of dialects. In order to assess the pace of 
variance changes with physical time and to calibrate the dating method we have used
historically attested events. Although the lack of documented historical events makes the 
direct calibration of the method difficult, we suggest (following \cite{Blanchard:2010a})
that variance evaluated over the Swadesh vocabulary proceeds approximately at the same 
pace uniformly for all human societies involved in trading and exchange. For calibrating 
the dating mechanism in \cite{Blanchard:2010a}, we 
have used the following four anchoring historical events (see \cite{Fouracre:2007}) for the 
Indo-European language family: i.) the last Celtic migration (to the Balkans and Asia Minor) 
(by 300 BCE); ii.) the division of the Roman Empire (by 500 CE); iii.) the migration of German 
tribes to the Danube River (by 100 CE); iv.) the establishment of the Avars Khaganate 
(by 590 CE) causing the spread of Slavic people. 
It is remarkable that all of the events mentioned uniformly indicate a very slow 
variance pace of a millionth per year, $t/\sigma^2=(1.367\pm 0.002)\cdot 10^6.$ 
This time-age ratio returns $t=1,353$ years if applied to the 
Malagasy dialects, suggesting that landing in Madagascar was around 650 CE.
This is in complete agreement with the prevalent opinion among scholars 
including the influential one of Adelaar \cite{Adelaar:2009}.

\section{The place of landing} 

In order to hypothetically infer the original center of dispersal of Malagasy variants,
we here use a variant of the method of \cite{Wichmann:2010a}. 
This method draws upon a well-known idea from biology \cite{Vavilov:1926} and linguistics 
\cite{Sapir:1916} that the homeland of a biological species or a language group corresponds 
to the current 
area of greatest diversity.
In \cite{Wichmann:2010a} this idea is transformed into quantifiable terms in the following way. 
For each language variant a diversity index is calculated as the average of the proportions 
between linguistic and geographical distances from the given language variant to each of 
the other language variants (cf. \cite{Wichmann:2010a} for more detail). 
The geographical distance is defined as the great-circle distance 
(i.e., as the crow flies) measured by angle radians.

In our work we adopted a variant of the method described in more detail in Appendix C.
The are two rationales for this variant, the first is that it
avoids technical and theoretical problems with pairs of dialects
which have a coinciding or a very close geographical location,
the second is that it uses the proper scaling proportions between lexical 
and geographical distances, extrapolated by linear regression.

The result of applying this method to Malagasy variants is that the best candidate for the 
homeland is the south-east coast where the three most diverse towns, i.e., Farafangana, 
Mahanoro and Ambovombe, are located, and where the surrounding towns are also highly diverse. 
The northern locations are the least diverse and they must have been settled last.

\begin{figure}
 \includegraphics[width=4.5truein,height=3.4truein,angle=0]{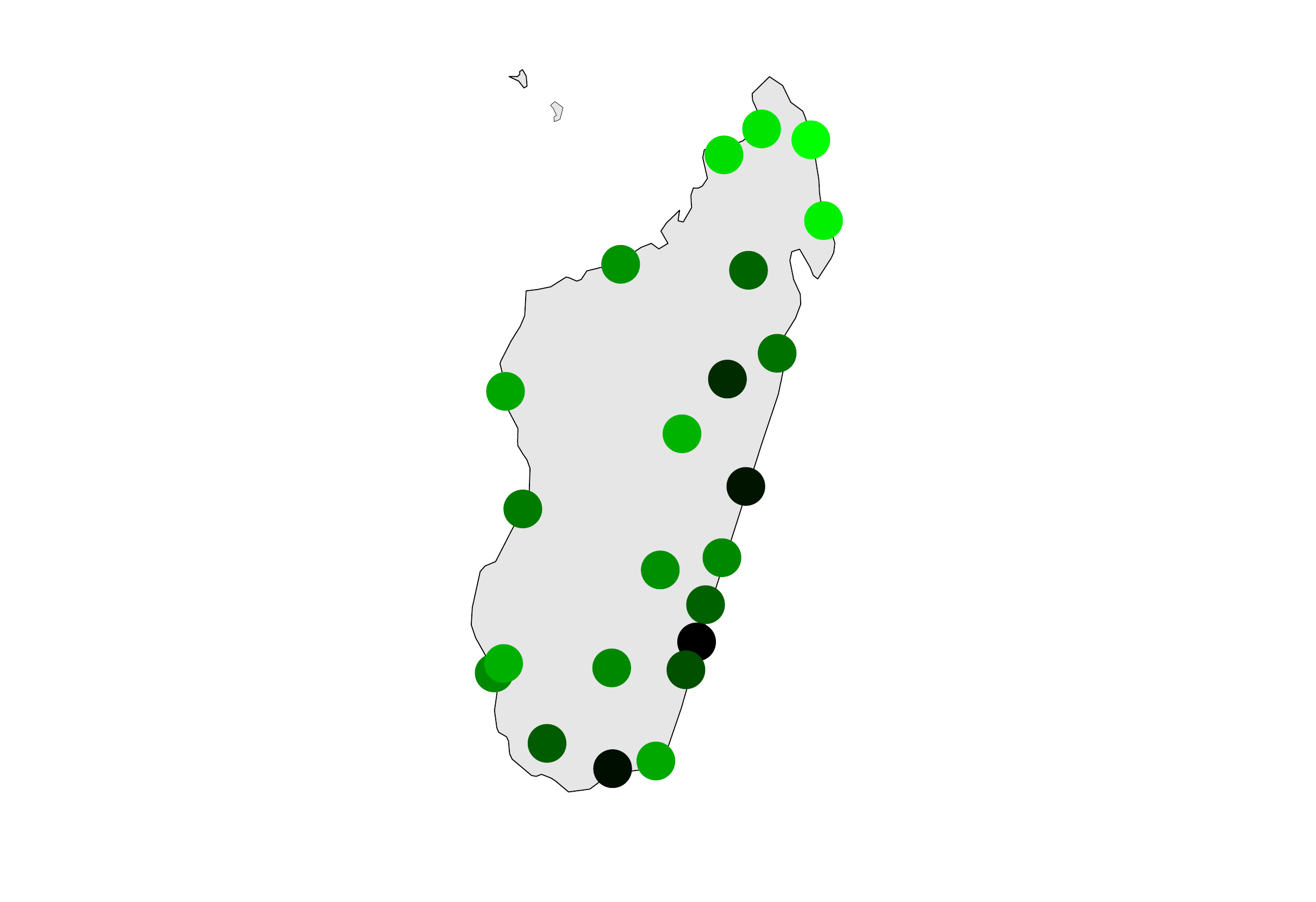}
 \caption{The homeland of Malagasy dialects as determined through diversity measures. 
The darkest-colored towns have the highest diversity values
while the light-colored the lowest. The most diverse area is the south-east coast where 
landing occurred, the less diverse area in the north, indicating that this area was settled last.}
\label{Fig_8}
\end{figure}

A convenient way of displaying the results on a map is shown in Fig. \ref{Fig_8},
where locations are indicated by means of 
circles with different gradations of the same color (green). The higher
the diversity index of a location is, the darker the color.
The figure suggest that the landing would have occurred somewhere between
Mahanoro (central part of the east cost) and Ambovobe (extreme 
south of the east coast), the most probable location being in the center of
this area, where Farafangana is situated.
Finally, we have checked that if the entire Greater Barito East group is considered, 
the homeland of Malagasy stays in the same place, but becomes secondary with respect 
to the southern Kalimantan homeland of the group.

The identification of a linguistic homeland for Malagasy on 
the south-east coast of 
Madagascar receives some independent support 
from unexpected kinds of evidence. 
According to \cite{Faublee:1983} there is an Indian Ocean current 
that connects Sumatra with 
Madagascar. When Mount Krakatoa exploded in 1883, pumice was washed ashore on 
Madagascar's east coast where the Mananjary River opens into the sea
(between Farafangana and Mahanoro). During World War II 
the same area saw the arrival of pieces of wreckage from ships sailing between 
Java and Sumatra that had been bombed by the Japanese air-force. 
The mouth of the Mananjary River is where the town of 
Manajary is presently located, 
and it is in the highly diverse south-east coast
as shown in Fig. \ref{Fig_8}. To enter the current that would eventually 
carry them to the east coast of Madagascar the ancestors of 
today's Malagasy people would 
likely have passed by the easily navigable Sunda strait. 

In his studies on the roots of Malagasy, Adelaar finds that the 
language has an important contingent of loanwords from Sulawesi (Buginese).
We also have compared Malagasy (and its dialects) with various 
Indonesian languages (Maanyan, Ngaju Dayak, Javanese, Iban, 
Banjarese, Bahasa Indonesia Malay, Manguidanaon, Maranao, Makassar, Buginese). 
While we unsurprisingly find that Maanyan is the closest language, 
we also find that
Buginese is the third closest one (see also \cite{Petroni:2008}).
The similarity with Buginese appears to be a further argument in 
favor of the southern 
path through the Sunda strait to Madagascar. If the correct scenario 
is one of Malay sailors 
recruiting a crew of other Austronesians and if the latter were 
recruited in Kalimantan and, to a limited extent, in Sulawesi,
then the settlers likely crossed this strait before starting their 
navigation in the open waters.

\begin{figure}
 \includegraphics[width=3.5truein,height=3.1truein,angle=0]{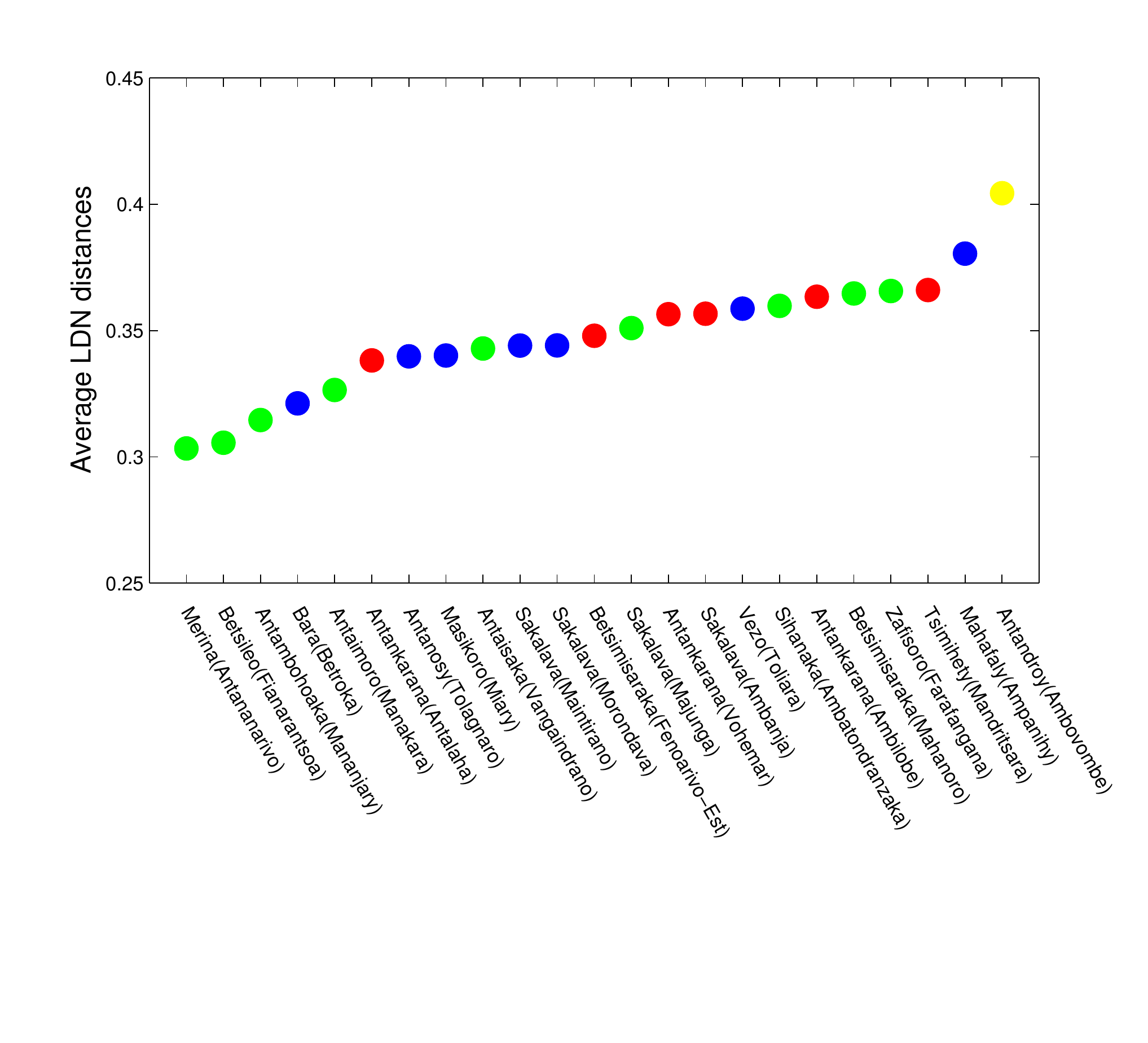}
 \caption{Average LDN distance of each of the dialects from all the others.   
Colors are chosen according to Figs. \ref{Fig_2}-\ref{Fig_3}.
Highland dialects (Antananarivo, Fianarantsoa and
Betroka) together with south-east coast dialects 
(Mananjary and Manakara) show the smallest average distance.}
\label{Fig_9}
\end{figure}

As a further confirmation of this analysis, we also computed the average
LDN distance from each dialect to all the others (see Fig.~\ref{Fig_9}).

Antandroy has the largest average distance, confirming that it is 
the overall most deviant variant (something which is also commonly 
pointed out by other Malagasy speakers).  
We further note that the smallest average distance is for Merina
(official language), Betsileo and Bara, which are all spoken 
in the highlands. 
The fact that Merina has the smallest average distance is 
possibly partially explained by the fact that this variant is the 
official one. However, as we will show later by means of a comparison 
of Malagasy dialects with Malay and Maanyan, this cannot 
be the only explanation.
More interestingly we remark that the Antambohoaka and Antaimoro variants, 
which are spoken in Mananjary and Manakara, also have a very small
average distance from the other dialects. Both dialects
are spoken in the south-east coast of Madagascar in a 
relatively isolated position and, therefore, this is further evidence for 
south-east as the homeland of the Malagasy language
and, likely, as the location of the first settlement.

%\begin{figure*}[ht]\centering 
% Using \begin{figure*} makes the figure as large as the entire page
%\includegraphics[width=\linewidth]{tabledist.pdf}
%\caption{A large .jpg figure}
%\label{ambatomilo}
%\end{figure*}

\section{Dialects, Malay and Maanyan}

The classification of Malagasy (together with all its dialects) among the 
Greater Barito East languages of Kalimantan as well as the particularly close 
relationship with Maanyan 
established in \cite{Dahl:1951} is beyond doubt.
However, Malagasy also underwent influences from other Indonesian languages 
such as Malay, Ngaju, Javanese, south-Sulawesi and south-Philippines 
languages \cite{Adelaar:1995a, Adelaar:2009}.

The main open problem concerning Malagasy is to determine the
composition of the population which settled the island.
Adelaar writes : {\it Malay influence persisted for several
centuries after the migration. But, except for this Malay influence, most
influence on Malagasy from other Indonesian languages seems to be 
pre-migratory. (...) I also believe it possible that the
early migrants from south-east Asia came not exclusively from the 
south-east Barito area, in fact, that south-east Barito speakers may 
not even have constituted a majority among these migrants, but rather formed a
nuclear group which was later reinforced by south-east Asian migrants
with a possibly different linguistic and cultural background (and, of course,
by African migrants).
Whatever view one may hold on how the early Malagasy were
influenced by other Indonesians, it seems necessary that we at least
develop a more cosmopolitan view on the Indonesian origins of the
Malagasy. A south-east Barito origin is beyond dispute, but this is of
course only one aspect of what Malagasy dialects and cultures reflect
today. Later influences were manifold, and some of these influences,
African as well as Indonesian, were so strong that they have molded the
Malagasy language and culture in all its variety into something new,
something for the analysis of which a south-east Barito origin has become a
factor of little explanatory value.}
 
\begin{figure}
 \includegraphics[width=3.5truein,height=3.5truein,angle=0]{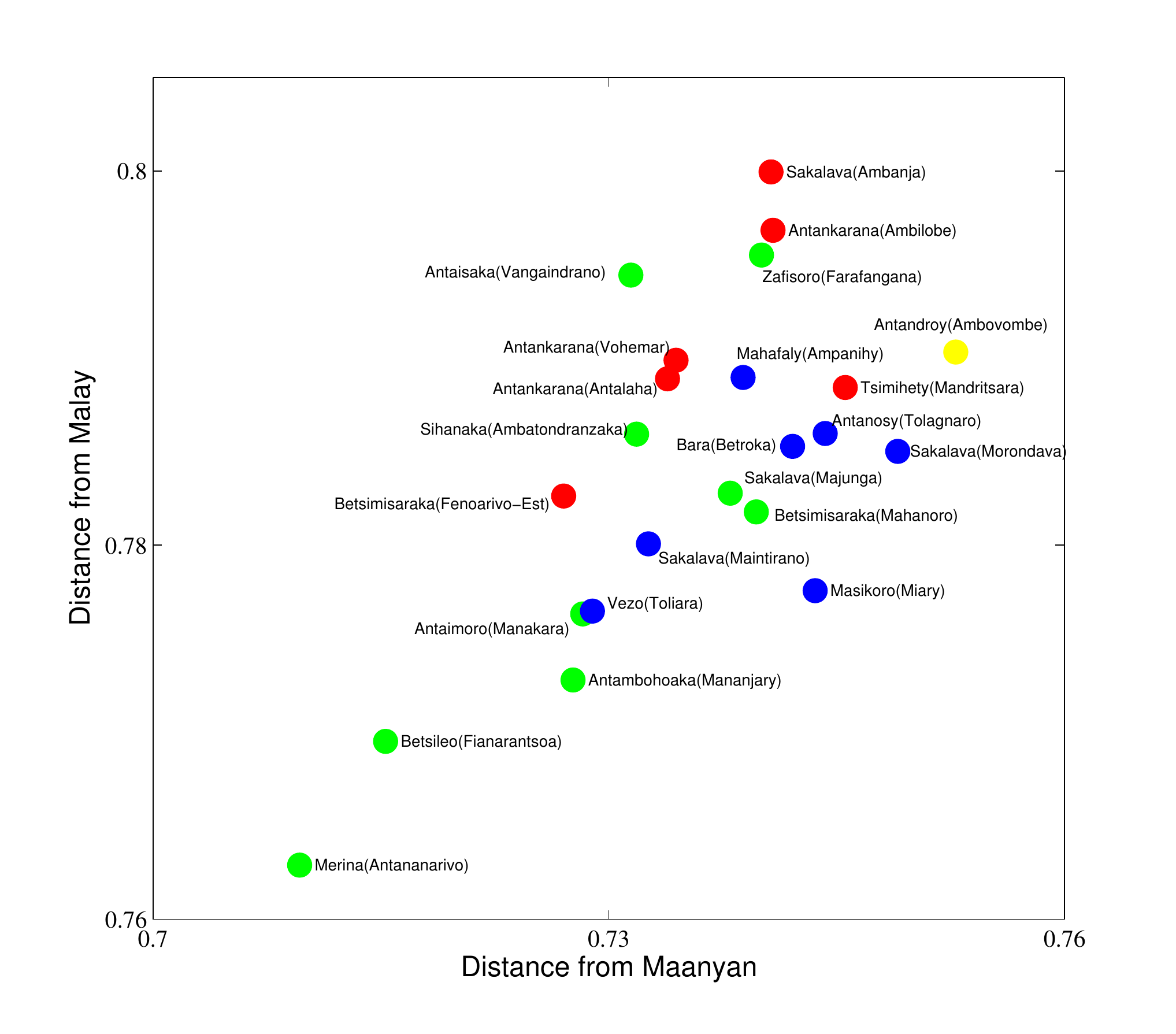}
 \caption{Lexical distances of Malagasy dialects from Malay and Maanyan. 
Colors are chosen according to Figs. \ref{Fig_2}-\ref{Fig_3}.
Highland dialects together with south-east coast Mananjary and Manakara 
dialects show the smallest distance from both the two Indonesian languages.}
\label{Fig_10}
\end{figure}

In order to clarify the problem raised by Adelaar, it is
necessary to understand the Malagasy relationships with other Indonesian
languages (and possibly African ones). 
The fact that the use of some words is limited to one or more dialects
was already taken into account in previous studies. For example it is known
that the word {\it alika} which refers to {\it dog} in Merina 
(the official variant) is replaced by the word {\it amboa} of Bantu
origin in most dialects.
Nevertheless, the study of Malagasy dialects in comparison with 
Indonesian languages is a still largely unexplored field of research.
Each dialect may provide pieces of information about the history the 
language,  eventually allowing us to for track the various linguistic 
influences experienced by Malagasy since the initial colonization of 
the island.
We know that Maanyan is one of the closest Indonesian language to Malagasy while
we also know that Malay somehow influenced Malagasy both before
and after colonization. In fact, there is also a possibility
that the first colonizers were a mixed equipage of Malay seafarers
and subordinates speaking en earlier form of Maanyan or a language closely to it, and that Malay influence 
continued for centuries (the first Malagasy alphabet
in Arabic characters was probably introduced by Muslim Malay seafarers).
For this reason we have computed the LDN distances of Malagasy dialects from Maanyan 
and Malay and we show them on the associated Cartesian plane (Fig. \ref{Fig_10}).

If we consider the 23 dialects together with Malay and Maanyan, not only do
we have to compute the 253 internal distances, but also we have to determine
the 23x2=56 distances of any of the dialects from the two Indonesian languages.
These new distances are displayed in Fig. \ref{Fig_10}.

First of all we observe, as expected, that the largest of the distances 
from Maanyan is smaller then the smallest of the distances from Malay.
This simply reflects the fact that Malagasy is first of all an East 
Barito language.
Then we also observe that Malagasy dialects seem to have almost the 
same relative composition. In fact, all the points in Fig. \ref{Fig_10} have 
almost the same {\it distance from Malay}/{\it distance from Maanyan} ratio.
This is a strong indication that the linguistic makeup
is substantially the same for all dialects
and, therefore, that they all originated by the same founding population 
of which they reflect the initial composition.
The conclusion is that the founding event
was likely a single one (\cite{Cox:2012})and subsequent immigration 
did not significantly alter the linguistic composition.

Indeed, looking more carefully, one can detect a little less Malay in the 
north since red circles have a larger ratio with 
respect all the others. This cannot a be a consequence of a larger 
African influence in the vocabulary due to the active trade with the continent 
and Comoros islands. In this case both the Maanyan and Malay 
component of the vocabulary would be affected. 
Instead, this may be the effect of Malay trading which, according to Adelaar, 
continued for several centuries after colonization. 
  
Noticeably, some dialects changed less with respect to the proto-language 
(Antananarivo, Fianarantsoa, Manajary, Manakara), in fact, their distances 
both from Maanyan and Malay are smaller then those of the other dialects. 
This is probably the most relevant phenomenon, and we 
underline that the variants which 
are less distant on average with respect to the other dialects 
(Fig. \ref{Fig_9}) are also less distant with respect to Malay and Maanyan (Fig. \ref{Fig_10}).
Therefore, the fact that Merina is closer to the other dialects cannot
be merely justified by the fact that it is the official variant.

We have checked whether the picture which emerges from Fig. \ref{Fig_10}
is confirmed by comparing with other related Indonesian languages.
The result is positive, and in particular the dialects of Manajary,
Manakara, Antananarivo and Fianarantsoa seem to be closer to 
most of the Indonesian languages which we compare them to.
Note that  Manajary and Manakara are both in the previously identified 
landing area on the south-east coast
while Antananarivo and Fianarantsoa are in the central highlands of the island.
This suggests a scenario according to which there was a migration 
to the highlands of Madagascar (Betsileo and Imerina regions)
shortly after the landing on the south-east coast (Manakara, Manajary).

In conclusion, both average distances in Fig. \ref{Fig_9} and distances 
from related Indonesian language (Fig. \ref{Fig_10}) point to the south-east coast
as the area of the first settlement.
This is the same indication which comes from the fact that linguistic 
diversity is higher in that region. 

Finally, we remark that the Antandroy variant (Ambovombe-be) 
is the most distant from Maanyan and 
among the most distant dialects from Malay, again showing itself to be the most 
deviant dialect. It is not clear whether its divergent evolution was due to 
internal factors or to specific language contacts which are still to be identified.

\section{Certainties and mysteries}

All results presented in this talk rely on two main ingredients: a new dataset 
from 23 different  variants of the languages (plus Malay and Maanyan)
and an automated method to evaluate lexical distances.
Analyzing the distances through different types of phylogenetic algorithm
(NJ and UPGMA) as well as through a geometrical approach we find that 
all approaches converge on a result 
where dialects are classified into two main geographical 
subgroups: south-west vs. center-north-east.
An output of the geometric representation of the distribution of the dialects 
is a landing date of around 650 CE, in agreement with a view commonly held by students of Malagasy.
Furthermore, by means of a technique which is based on the calculation 
of differences in linguistic diversity, we propose that the south-east 
coast was the location were the first colonizers landed. 
This location also suggests that the path followed by the sailors 
went from Kalimantan, through the Sunda strait, and subsequently, 
along major oceanic currents, to Madagascar.
We also measured the distance of the Malagasy variants to other Indonesian languages 
and found that the dialects of Manajary, Manakara, Antananarivo and 
Fianarantsoa are noticeably closer to most of them as well as closer, 
on average, to the other variants of the language.
Manajary and Manakara are both in the identified landing area in the 
south-east coast which is therefore confirmed.
Antananarivo and Fianarantsoa are in the central highlands 
of Madagascar suggesting that 
landing was followed shortly after by a migration to the interior of the island.
A measure of the average distance of any single dialect with 
respect to all the others leads to the same conclusions.
Finally, comparison with Maanyan and Malay suggests a single colonization event.
\smallskip

Together with these certainties, there are still some mysteries 
concerning previous peopling of Madagascar (eventual inhabitants
before the founding event in 650 CE) and
ancestry (ethnic composition of Indonesian colons).

The island was almost surely inhabited before the arrival of Malagasy ancestors.
Malagasy mythology portrays a people, called the {\it Vazimba}, as the 
original inhabitants, and it is not clear whether they were part of a 
previous Austronesian expansion or a population
of a completely different origin (Bantu, Khoisan?). 
In the latter case it may be 
possible to track the aboriginal vocabulary in
the dialects. For example, the Mikea are the only hunter-gatherers 
in Madagascar, and it is unclear whether they are a relic of the
aboriginal pre-Indonesian population or just 'ordinary' Malagasy who
switched to a simpler economy for historical reasons.
If the first hypothesis is the correct one (see \cite{Blench:2010}),
they should show some residual aboriginal vocabulary in their dialect, 
and the same is expected for the neighboring populations of 
Vezo and Masikoro.

Despite the strong linguistic affinity, it seems that the Maanyan of southern Kalimantan
are not the primary source population of the Malagasy.
In fact, \cite{Kusuma:2016} evidenced that the Maanyan are characterized by a 
distinct, high frequency genomic component that is not found in the Malagasy. 
In contrast, they found that the Malagasy show strong genomic links to  a range of 
southern Kalimantan groups as the Banjar.
Moreover, the Indonesian makeup of Malagasy also extends 
to a range of insular southeast Asian groups.
In fact, both maternal and paternal DNA lineages suggest \cite{Kusuma:2015} that
Malagasy derive from multiple regional sources in Indonesia, 
which also include southern Sulawesi and the Lesser Sunda islands.
We have already discussed that Malagasy language also records this
multiple composition of the Indonesian colonizers \cite{Adelaar:1995a, Adelaar:2009}.
It may be possible that this still shadowed evidence
is more clearly encoded in some Malagasy variants which were still
not sufficiently considered.

These two mysteries, {\it Vazimba} and ancestry,
call for a new look at the Malagasy language,
not as a single entity, but as a constellation of variants 
whose histories are still to be fully understood. 

%------------------------------------------------

\section*{Acknowledgments} % The \section*{} command stops section numbering

I am deeply indebted with Emilienne Aim\'ee Razafindratema and
Joselin\`a Soafara N\'{e}r\'{e} for their invaluable
help in collecting data.

I am also indebted with Clement {\it Zazalahy}, Beatrice Rolla, 
Renato Magrin and Corto Maltese
for logistical support during my stay in Madagascar. 

Special thanks to Armando Neves for his kind invitation and for all
time he wasted to give me the possibility to visit once more the
UFMG.

\section*{Appendix A}

The lexical distance (as defined in  
\cite{Serva:2008}, \cite{Petroni:2008} and \cite{ Holman:2008}) between the two 
languages,  $l_i$ and $l_j$, is computed as the average of the normalized Levenshtein 
(edit) distance 
\cite{Levenshtein:1966} 
over the vocabulary of 200 items,
\begin{equation}
D\left(l_i,l_j\right)\,\,=\,\, \frac{1}{200}\sum_{\alpha=1}^{200}
\frac{\left\| w_i (\alpha),w_j (\alpha) \right\|}
{\max \left( \left| w_i (\alpha) \right|,\left| w_j (\alpha)\right|
\label{distance} 
\right)}.
\end{equation}

\noindent
where the item is indicated by $\alpha$,
 $\left\| w_i(\alpha),w_j(\alpha) \right\|$ is the standard Levenshtein distance 
between the words $w_i(\alpha)$ and $w_j(\alpha)$, 
and $|w_i(\alpha)|$ is the number of characters in the word $w_i(\alpha)$ . 
The sum runs over all the 200 different items of the Swadesh list.
Assuming that the number of languages (or dialects) 
to be compared is $N$, then the
distances $D(l_i,l_j)$  are the entries of a $N \times N$ symmetric matrix ${\bf D}$
(obviously $D(l_i,l_i)=0$).
The matrix can be found in \cite{Serva:2012a, Serva:2012b}.

\section*{Appendix B}

The lexical distance (\ref{distance}) between two languages, $l_i$ 
and $l_j$, can be interpreted as the average probability to distinguish them by a mismatch 
between two characters randomly chosen from the orthographic realizations of the vocabulary 
meanings. There are infinitely many matrices that match all the structure of ${\bf D}$, 
and therefore contain all the information about the relationships between languages, 
\cite{Blanchard:2010a}. It is remarkable that all these matrices are related to each 
other by means of a linear transformation,
\begin{equation}
\label{random_walk}
{\bf T }\,\,=\,\,
{\bf R}^{-1} {\bf D },\quad 
{\bf R}=\mathrm{diag}\left(
\sum_{k=1}^N D\left(l_1,l_k\right)\ldots
 \sum_{k=1}^N D\left(l_N,l_k\right)
\right)
\end{equation}
which can be interpreted as the generator of a random walk on the weighted undirected 
graph determined by the matrix of lexical distances ${\bf D}$ over the $N$ different languages 
\cite{Blanchard:2010a,Blanchard:2010b}. The
random walks defined by the transition matrix 
(\ref{random_walk}) describe the statistics of a sequential process of language 
classification. Namely, while the elements $T(l_i, l_j)$ of the matrix ${\bf T}$ evaluate 
the probability of successful differentiation of the language $l_i$ provided the 
language $l_j$ has been identified certainly, the elements of 
the squared matrix ${\bf T}^2$, 
ascertain the successful differentiation of the language $l_i$ from $l_j$ through an 
intermediate language, the elements of the matrix ${\bf T}^3$ give the probabilities to 
differentiate the language through two intermediate steps, and so on. 
The whole host of complex and indirect relationships between orthographic 
representations of the vocabulary meanings encoded in the matrix of lexical 
distances (\ref{distance}) is uncovered by the von Neuman series estimating the 
characteristic time of successful classification for any two languages in the database 
over a language family,

\begin{equation}
\label{kernel}
{\bf J}\,\,=\,\,
\lim_{n\to\infty}
\sum_{k=0}^n {\bf T}^n\,\,=\,\, \frac 1{1-{\bf T}}.
\end{equation}

The last equality in (\ref{kernel}) is understood as the group generalized inverse 
(Blanchard:2010b) being a symmetric, positive semi-definite matrix which plays the 
essentially same role for the SCA, as the covariance matrix does for the usual PCA analysis. 
The standard goal of a component analysis (minimization of the data redundancy quantified by 
the off-diagonal elements of the kernel matrix) is readily achieved by solving an eigenvalue 
problem for the matrix ${\bf J}$. Each column vector $q_k$, which determines a direction 
where $\bf J$ acts as a simple rescaling, ${\bf J}q_k = \lambda_kq_k$, with some real 
eigenvalue $\lambda_k = 0$, is associated to the virtually independent trait in the matrix 
of lexical distances ${\bf D}$. Independent components 
$\left\{q_k\right\}$, $k = 1,\ldots N$, define an orthonormal basis in $\mathbb{R}^N$ 
which specifies each language $l_i$ by $N$ numerical coordinates, 
$l_i \to \left(q_{1,i}, q_{2,i},\ldots q_{N,i}\right)$. 
Languages that cast in the same mold in accordance with the $N$ individual 
data features are revealed by geometric proximity in Euclidean space spanned by the 
eigenvectors $\left\{q_k\right\}$ that might be either exploited visually, or accounted 
analytically. The rank-ordering of data traits $\left\{q_k\right\}$, in accordance to their 
eigenvalues, $\lambda_0 =\lambda_1 <\lambda_2 =\ldots= \lambda_N$, provides us with the
 natural geometric framework for dimensionality reduction. At variance with the standard 
PCA analysis \cite{Jolliffe:2002}, where the largest eigenvalues of the covariance matrix 
are used in order to identify the principal components, while building language taxonomy, 
we are interested in detecting the groups of the most similar languages, with respect to 
the selected group of features. The components of maximal similarity are identified with the 
eigenvectors belonging to the smallest non-trivial eigenvalues. Since the minimal 
eigenvalue $\lambda_1 = 0$ corresponds to the vector of stationary distribution of random 
walks and thus contains no information about components, we have used the three consecutive 
components $\left(q_{2,i}, q_{3,i},q_{4,i}\right)$ as the three Cartesian coordinates of a 
language $l_i$ in order to build a three-dimensional geometric representation 
of language taxonomy. Points symbolizing different languages in space of the three major 
data traits are contiguous if the orthographic representations of the vocabulary meanings 
in these languages are similar.

\section*{Appendix C}

\noindent 
The lexical distance $D(l_i,l_j)$ between two dialects $l_i$ and $l_j$
was previously defined; their geographical distance $\Delta(l_i,l_j)$  can be simply
defined as the distance between the two locations where the dialects were collected.
There are different possible measure units for $\Delta(l_i,l_j)$.
We simply use the great-circle angle (the angle that the two location form with the center 
of the earth).

It is reasonable to assume, in general, that larger geographical distances
correspond to larger lexical distances and vice-versa
For this reason in \cite{Wichmann:2010a} the diversity was measured 
as the average of the ratios between lexical and geographical distance.
This definition implicitly assumes that lexical distances vanish
when geographical distances equal 0. Nevertheless, different dialects 
are often spoken at the same locations, separated by negligible geographical
distances. For this reason, and because a zero denominator in the division involving
geographical distances would cause some diversity indexes to become infinite,
\cite{Wichmann:2010a} arbitrarily added a constant of .01 km to all distances. 

Here we similarly add a constant, but one whose value is better motivated.
We plotted all the $\frac {23 \times 22}{2}=253$ points 
$\Delta(l_i,l_j), \, D(l_i,l_j)$ in a bi-dimensional space
and verified that the pattern is compatible with a linear shape
in the domain of small geographical distances.
Linear regression of the $20\%$ of points with smaller 
geographical distances  gives the interpolating straight line $D=a+b\Delta$ with 
$a=0.22$ and $b=0.04$.
The results indicates that a lexical distance of $0.22$ is 
expected between two variants of a language spoken in coinciding locations. 

The choice of constants $a$ and $b$ by linear regression
assures that the ratio between $D(l_i,l_j)$ and $a + b\Delta(l_i,l_j)$ is around 1
for any pair of dialects  $l_i$ and $l_j$. 
A large value of the ratio corresponds to a pair of variants which are lexically 
more distant and vice-versa.
It is straightforward to define the diversity of a dialect as 

\begin{equation}
V(l_i) \,\, = \,\, \frac{1}{22} \sum_{j \neq i}
\frac{D(l_i,l_j)}{a + b\Delta(l_i,l_j)}
\label{diversity}
\end{equation}
in this way, locations with high diversity will be characterized by a a larger $V(l_i)$,
while locations with low diversity will have a smaller one.

Notice that the above definition coincides with the one in \cite{Wichmann:2010a},
the main difference being that instead of an arbitrary value of $a$ we obtain
it through the output of linear regression. 

The diversities (in a decreasing order), computed with (\ref{diversity}), are the following:
Zafisoro (Farafangana): 1.00,  
Betsimisaraka (Mahanoro): 0.98,
Antandroy (Ambovombe): 0.98, 
Sihanaka (Ambatontrazaka): 0.95,
Antaisaka (Vangaindrano): 0.92,  
Mahafaly (Ampanihy): 0.90, 
Tsimihety (Mandritsara): 0.90, 
Antaimoro (Manakara): 0.90, 
Betsimisaraka (Fenoarivo-Est): 0.88, 
Sakalava (Morondava): 0.87, 
Antambohoaka (Mananjary): 0.86,
Vezo (Toliara): 0.86, 
Bara (Betroka): 0.86,
Sakalava (Majunga): 0.85,
Betsileo (Fianarantsoa): 0.85,
Antanosy (Tolagnaro): 0.83, 
Sakalava (Maintirano): 0.83, 
Masikoro (Miary): 0.82,
Merina (Antananarivo): 0.82, 
Sakala-Sakalava (Ambanja): 0.77, 
Antankarana (Ambilobe): 0.77,
Antankarana (Antalaha): 0.76,
Antankarana (Vohemar): 0.74.

%--------------------------------------------------------------------------------
%	REFERENCES LIST
%--------------------------------------------------------------------------------

%-------------------------------------------------------------------------------

\end{document}